\documentclass[runningheads]{llncs}
\usepackage{multirow}
\usepackage[T1]{fontenc}
\usepackage[utf8]{inputenc}
\usepackage{graphicx}
\usepackage{wrapfig}
\usepackage{subcaption}
\usepackage{xcolor}
\usepackage[most]{tcolorbox}
\usepackage[draft]{minted}
\makeatletter
\def\PYG@reset{\let\PYG@it=\relax \let\PYG@bf=\relax%
    \let\PYG@ul=\relax \let\PYG@tc=\relax%
    \let\PYG@bc=\relax \let\PYG@ff=\relax}
\def\PYG@tok#1{\csname PYG@tok@#1\endcsname}
\def\PYG@toks#1+{\ifx\relax#1\empty\else%
    \PYG@tok{#1}\expandafter\PYG@toks\fi}
\def\PYG@do#1{\PYG@bc{\PYG@tc{\PYG@ul{%
    \PYG@it{\PYG@bf{\PYG@ff{#1}}}}}}}
\def\PYG#1#2{\PYG@reset\PYG@toks#1+\relax+\PYG@do{#2}}

\@namedef{PYG@tok@w}{\def\PYG@tc##1{\textcolor[rgb]{0.73,0.73,0.73}{##1}}}
\@namedef{PYG@tok@c}{\let\PYG@it=\textit\def\PYG@tc##1{\textcolor[rgb]{0.24,0.48,0.48}{##1}}}
\@namedef{PYG@tok@cp}{\def\PYG@tc##1{\textcolor[rgb]{0.61,0.40,0.00}{##1}}}
\@namedef{PYG@tok@k}{\let\PYG@bf=\textbf\def\PYG@tc##1{\textcolor[rgb]{0.00,0.50,0.00}{##1}}}
\@namedef{PYG@tok@kp}{\def\PYG@tc##1{\textcolor[rgb]{0.00,0.50,0.00}{##1}}}
\@namedef{PYG@tok@kt}{\def\PYG@tc##1{\textcolor[rgb]{0.69,0.00,0.25}{##1}}}
\@namedef{PYG@tok@o}{\def\PYG@tc##1{\textcolor[rgb]{0.40,0.40,0.40}{##1}}}
\@namedef{PYG@tok@ow}{\let\PYG@bf=\textbf\def\PYG@tc##1{\textcolor[rgb]{0.67,0.13,1.00}{##1}}}
\@namedef{PYG@tok@nb}{\def\PYG@tc##1{\textcolor[rgb]{0.00,0.50,0.00}{##1}}}
\@namedef{PYG@tok@nf}{\def\PYG@tc##1{\textcolor[rgb]{0.00,0.00,1.00}{##1}}}
\@namedef{PYG@tok@nc}{\let\PYG@bf=\textbf\def\PYG@tc##1{\textcolor[rgb]{0.00,0.00,1.00}{##1}}}
\@namedef{PYG@tok@nn}{\let\PYG@bf=\textbf\def\PYG@tc##1{\textcolor[rgb]{0.00,0.00,1.00}{##1}}}
\@namedef{PYG@tok@ne}{\let\PYG@bf=\textbf\def\PYG@tc##1{\textcolor[rgb]{0.80,0.25,0.22}{##1}}}
\@namedef{PYG@tok@nv}{\def\PYG@tc##1{\textcolor[rgb]{0.10,0.09,0.49}{##1}}}
\@namedef{PYG@tok@no}{\def\PYG@tc##1{\textcolor[rgb]{0.53,0.00,0.00}{##1}}}
\@namedef{PYG@tok@nl}{\def\PYG@tc##1{\textcolor[rgb]{0.46,0.46,0.00}{##1}}}
\@namedef{PYG@tok@ni}{\let\PYG@bf=\textbf\def\PYG@tc##1{\textcolor[rgb]{0.44,0.44,0.44}{##1}}}
\@namedef{PYG@tok@na}{\def\PYG@tc##1{\textcolor[rgb]{0.41,0.47,0.13}{##1}}}
\@namedef{PYG@tok@nt}{\let\PYG@bf=\textbf\def\PYG@tc##1{\textcolor[rgb]{0.00,0.50,0.00}{##1}}}
\@namedef{PYG@tok@nd}{\def\PYG@tc##1{\textcolor[rgb]{0.67,0.13,1.00}{##1}}}
\@namedef{PYG@tok@s}{\def\PYG@tc##1{\textcolor[rgb]{0.73,0.13,0.13}{##1}}}
\@namedef{PYG@tok@sd}{\let\PYG@it=\textit\def\PYG@tc##1{\textcolor[rgb]{0.73,0.13,0.13}{##1}}}
\@namedef{PYG@tok@si}{\let\PYG@bf=\textbf\def\PYG@tc##1{\textcolor[rgb]{0.64,0.35,0.47}{##1}}}
\@namedef{PYG@tok@se}{\let\PYG@bf=\textbf\def\PYG@tc##1{\textcolor[rgb]{0.67,0.36,0.12}{##1}}}
\@namedef{PYG@tok@sr}{\def\PYG@tc##1{\textcolor[rgb]{0.64,0.35,0.47}{##1}}}
\@namedef{PYG@tok@ss}{\def\PYG@tc##1{\textcolor[rgb]{0.10,0.09,0.49}{##1}}}
\@namedef{PYG@tok@sx}{\def\PYG@tc##1{\textcolor[rgb]{0.00,0.50,0.00}{##1}}}
\@namedef{PYG@tok@m}{\def\PYG@tc##1{\textcolor[rgb]{0.40,0.40,0.40}{##1}}}
\@namedef{PYG@tok@gh}{\let\PYG@bf=\textbf\def\PYG@tc##1{\textcolor[rgb]{0.00,0.00,0.50}{##1}}}
\@namedef{PYG@tok@gu}{\let\PYG@bf=\textbf\def\PYG@tc##1{\textcolor[rgb]{0.50,0.00,0.50}{##1}}}
\@namedef{PYG@tok@gd}{\def\PYG@tc##1{\textcolor[rgb]{0.63,0.00,0.00}{##1}}}
\@namedef{PYG@tok@gi}{\def\PYG@tc##1{\textcolor[rgb]{0.00,0.52,0.00}{##1}}}
\@namedef{PYG@tok@gr}{\def\PYG@tc##1{\textcolor[rgb]{0.89,0.00,0.00}{##1}}}
\@namedef{PYG@tok@ge}{\let\PYG@it=\textit}
\@namedef{PYG@tok@gs}{\let\PYG@bf=\textbf}
\@namedef{PYG@tok@gp}{\let\PYG@bf=\textbf\def\PYG@tc##1{\textcolor[rgb]{0.00,0.00,0.50}{##1}}}
\@namedef{PYG@tok@go}{\def\PYG@tc##1{\textcolor[rgb]{0.44,0.44,0.44}{##1}}}
\@namedef{PYG@tok@gt}{\def\PYG@tc##1{\textcolor[rgb]{0.00,0.27,0.87}{##1}}}
\@namedef{PYG@tok@err}{\def\PYG@bc##1{{\setlength{\fboxsep}{\string -\fboxrule}\fcolorbox[rgb]{1.00,0.00,0.00}{1,1,1}{\strut ##1}}}}
\@namedef{PYG@tok@kc}{\let\PYG@bf=\textbf\def\PYG@tc##1{\textcolor[rgb]{0.00,0.50,0.00}{##1}}}
\@namedef{PYG@tok@kd}{\let\PYG@bf=\textbf\def\PYG@tc##1{\textcolor[rgb]{0.00,0.50,0.00}{##1}}}
\@namedef{PYG@tok@kn}{\let\PYG@bf=\textbf\def\PYG@tc##1{\textcolor[rgb]{0.00,0.50,0.00}{##1}}}
\@namedef{PYG@tok@kr}{\let\PYG@bf=\textbf\def\PYG@tc##1{\textcolor[rgb]{0.00,0.50,0.00}{##1}}}
\@namedef{PYG@tok@bp}{\def\PYG@tc##1{\textcolor[rgb]{0.00,0.50,0.00}{##1}}}
\@namedef{PYG@tok@fm}{\def\PYG@tc##1{\textcolor[rgb]{0.00,0.00,1.00}{##1}}}
\@namedef{PYG@tok@vc}{\def\PYG@tc##1{\textcolor[rgb]{0.10,0.09,0.49}{##1}}}
\@namedef{PYG@tok@vg}{\def\PYG@tc##1{\textcolor[rgb]{0.10,0.09,0.49}{##1}}}
\@namedef{PYG@tok@vi}{\def\PYG@tc##1{\textcolor[rgb]{0.10,0.09,0.49}{##1}}}
\@namedef{PYG@tok@vm}{\def\PYG@tc##1{\textcolor[rgb]{0.10,0.09,0.49}{##1}}}
\@namedef{PYG@tok@sa}{\def\PYG@tc##1{\textcolor[rgb]{0.73,0.13,0.13}{##1}}}
\@namedef{PYG@tok@sb}{\def\PYG@tc##1{\textcolor[rgb]{0.73,0.13,0.13}{##1}}}
\@namedef{PYG@tok@sc}{\def\PYG@tc##1{\textcolor[rgb]{0.73,0.13,0.13}{##1}}}
\@namedef{PYG@tok@dl}{\def\PYG@tc##1{\textcolor[rgb]{0.73,0.13,0.13}{##1}}}
\@namedef{PYG@tok@s2}{\def\PYG@tc##1{\textcolor[rgb]{0.73,0.13,0.13}{##1}}}
\@namedef{PYG@tok@sh}{\def\PYG@tc##1{\textcolor[rgb]{0.73,0.13,0.13}{##1}}}
\@namedef{PYG@tok@s1}{\def\PYG@tc##1{\textcolor[rgb]{0.73,0.13,0.13}{##1}}}
\@namedef{PYG@tok@mb}{\def\PYG@tc##1{\textcolor[rgb]{0.40,0.40,0.40}{##1}}}
\@namedef{PYG@tok@mf}{\def\PYG@tc##1{\textcolor[rgb]{0.40,0.40,0.40}{##1}}}
\@namedef{PYG@tok@mh}{\def\PYG@tc##1{\textcolor[rgb]{0.40,0.40,0.40}{##1}}}
\@namedef{PYG@tok@mi}{\def\PYG@tc##1{\textcolor[rgb]{0.40,0.40,0.40}{##1}}}
\@namedef{PYG@tok@il}{\def\PYG@tc##1{\textcolor[rgb]{0.40,0.40,0.40}{##1}}}
\@namedef{PYG@tok@mo}{\def\PYG@tc##1{\textcolor[rgb]{0.40,0.40,0.40}{##1}}}
\@namedef{PYG@tok@ch}{\let\PYG@it=\textit\def\PYG@tc##1{\textcolor[rgb]{0.24,0.48,0.48}{##1}}}
\@namedef{PYG@tok@cm}{\let\PYG@it=\textit\def\PYG@tc##1{\textcolor[rgb]{0.24,0.48,0.48}{##1}}}
\@namedef{PYG@tok@cpf}{\let\PYG@it=\textit\def\PYG@tc##1{\textcolor[rgb]{0.24,0.48,0.48}{##1}}}
\@namedef{PYG@tok@c1}{\let\PYG@it=\textit\def\PYG@tc##1{\textcolor[rgb]{0.24,0.48,0.48}{##1}}}
\@namedef{PYG@tok@cs}{\let\PYG@it=\textit\def\PYG@tc##1{\textcolor[rgb]{0.24,0.48,0.48}{##1}}}

\def\PYGZus{\char`\_}
\def\PYGZob{\char`\{}
\def\PYGZcb{\char`\}}

\def\PYGZlt{\char`\<}
\def\PYGZgt{\char`\>}

\def\PYGZhy{\char`\-}

\makeatother

\usepackage{tikz}
\usetikzlibrary{arrows.meta}
\usetikzlibrary{positioning, patterns, calc, shapes, arrows, shapes.misc}
\usetikzlibrary{decorations.pathreplacing, shapes.multipart}
\tikzstyle{neuron}=[draw,circle,thick,inner sep=0, minimum size = 20]
\tikzstyle{cfgnode}=[draw,circle,thick,inner sep=5,
minimum width = 30, blue, fill=blue!5, text=black, align=left]
\tikzstyle{inactivation} = [cross out, draw=red, line width=0.75mm,
minimum size = 10pt, inner sep=0pt, outer sep=0pt]
\tikzstyle{activation} = [circle, draw=olive, line width=0.75mm,
minimum size = 10pt]
\tikzstyle{wv} = [line width=0.75mm, midway, above]
\tikzstyle{modulesingle} = [rectangle, draw, thick, align=center,
inner sep=0.2cm,
orange, text=black, fill = orange!20, rounded corners]
\usetikzlibrary{calc}
\usepackage{relsize}

\tikzset{fontscale/.style = {font=\relsize{#1}}
    }

\usepackage{booktabs}
\usepackage{amsmath}
\usepackage{amsfonts}
\usepackage{amssymb}
\usepackage{amssymb}
\usepackage{subcaption}
\usepackage{mathtools}
\usepackage{comment}
\usepackage{url}
\usepackage{xspace}
\usepackage{listings}
\usepackage{bbm}
\newcommand{\hypothesis}{\mathcal{H}}
\newcommand{\problemSpace}{\mathcal{P}}
\newcommand{\resultSpace}{\mathcal{R}}

\newcommand{\system}{s}
\newcommand{\systemProperty}{\Psi}
\newcommand{\solutionProperty}{\Phi}
\newcommand{\networkProperty}{\Xi}

\newcommand{\vnnc}{VNN-COMP\xspace}
\newcommand{\vnnl}{VNN-LIB\xspace}
\newcommand{\smtl}{SMT-LIB\xspace}
\newcommand{\onnx}{ONNX\xspace}

\newcommand{\caisar}{CAISAR\xspace}
\newcommand{\R}{\mathbb{R}}

\usepackage{caption}
\usepackage{todonotes}

\begin{document}
\title{Neural Network Verification is\\ a Programming Language Challenge }
\titlerunning{NN Verification is a PL Challenge}
%
\author{Lucas C. Cordeiro\inst{1} \and Matthew L. Daggitt\inst{2} \and Julien Girard-Satabin\inst{3} \and Omri Isac\inst{4} \and Taylor T. Johnson\inst{5} \and Guy Katz\inst{4} \and Ekaterina Komendantskaya\inst{6},\inst{7} \and Augustin Lemesle\inst{3} \and Edoardo Manino\inst{1} \and Artjoms {\v{S}}inkarovs\inst{6} \and Haoze Wu\inst{8}}

\institute{University of Manchester, UK \and University of Western Australia, Australia \and Atomic Energy and Alternative Energies Commission, France \and Hebrew University of Jerusalem, Israel \and Vanderbilt University, USA \and Southampton University, UK \and Heriot-Watt Univerwsity, UK \and Amherst College, USA }

\authorrunning{L. C. Cordeiro et al.}


%
%
\maketitle              
\begin{abstract}
Neural network verification is a new and rapidly developing field of research.
So far, the main priority has been establishing efficient verification algorithms and tools, while proper support from the programming language perspective has been considered secondary or unimportant. Yet, there is mounting evidence that insights from the programming language community may make a difference in the future development of this domain. In this paper, we formulate neural network verification challenges as programming language challenges and suggest possible future solutions. 

\keywords{Neural Networks \and Verification  \and Domain Specific Languages.}
\end{abstract}
%
%

%
\section{Introduction}
\label{sec:intro}


Traditionally, statistical machine learning has distinguished its methods from ``algorithm-driven'' programming: the consensus has been that machine learning is deployed when there is example input-output data but no general algorithm for computing outputs from inputs. Thus, neural networks are commonly seen as programs that emerge from data via training, without direct human guidance on how to perform the computation. This unfortunate dichotomy has led to a divide between programming language and machine learning research that is still awaiting resolution. 

The first hint that this dichotomy is not as fundamental as was thought came from the machine learning community itself. The famous paper by Szegedy et al.~\cite{szegedy2014intriguing} pointed out the ``intriguing'' problem that even the most accurate neural networks fail to satisfy the property of \emph{robustness}, i.e. small perturbations of their inputs should result in small changes to their output. Szegedy's key example concerned imperceptible perturbations of pixels in an image that can sway the neural network's classification decisions. This lack of robustness can have safety and security implications: for example, an autonomous car's vision unit may fail to recognise pedestrians on the road. For that reason, the problem attracted significant attention~\cite{Carlini19} but remains unresolved to this day.
Partial solutions often deploy methods of \emph{adversarial training} --- i.e., training based on computing \emph{adversarial attacks} --- which augment the training set with the worst-case perturbations of the input data points with respect to the output loss of the neural network~\cite{KM18}.

The robustness of neural networks actually yields a formal specification~\cite{CKDKKAE22}. Given a neural network $f: \R^m \to \R^n$, $f$ is \emph{robust around $\hat{x} \in \R^m$}, if 
 \begin{align}
 \label{robustness}  
 \forall x, \| \hat{x} - x\| \leq \epsilon  \implies  \| f(\hat{x}) - f(x)\| \leq \delta ,   
 \end{align}
 
\noindent where $\epsilon, \delta \in \R$ are small constants and $\|.\|$ computes a vector distance. From the programming language perspective, robustness can be seen as a refinement type that refines input and output types of $f$, cf.~\cite{KKKAA20}.
%
At the same time, robustness is an example of a \emph{desirable property} that neural networks cannot
learn from data alone: note the quantification over vectors $x$ that do not belong to the data set. 
This challenges the classical dichotomy between algorithm-driven and data-driven programming, demonstrating the inevitability of property specification in both cases.

\begin{figure}[t!]
\includegraphics[width=\textwidth]{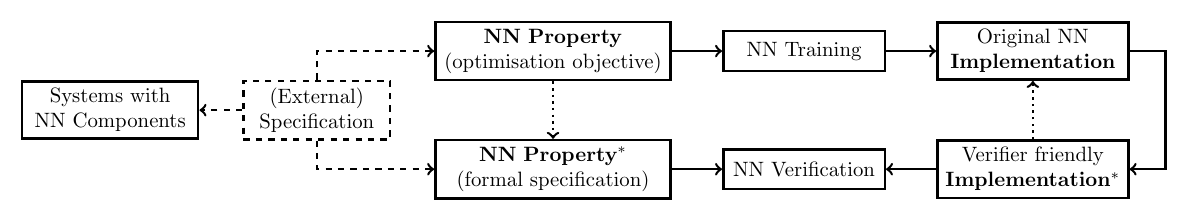}
\caption{\footnotesize{\emph{Schematic representation of the state of the art in training and verifying neural networks for properties. Solid lines denote methods widely accepted by the research communities, dashed lines mean ``some experimental prototypes exist'', dotted arrows mean the connection is desired but not established.}}}\label{fig:CV}
\end{figure}

Against this background, both the machine learning and verification communities proposed several useful methods of training for, or respectively verifying, \emph{certain properties}\footnote{\scriptsize{We deliberately use the term ``properties" rather than ``specifications" here, as the latter means the presence of a sufficiently general specification language.}}.
Fig.~\ref{fig:CV} depicts these two groups of methods as two parallel pipelines.
At the top, we include all adversarial training methods~\cite{KM18} that were generalised to account for arbitrary optimisation objectives, given a property informally expressed in (a fragment of)  first-order logic~\cite{FischerBDGZV19,ijcai2022p767}. 
At the bottom, we include the verification pipeline which is supported by more than a dozen neural network verifiers,
such as Marabou~\cite{katz2019marabou,Wu24}, $\alpha\beta$-CROWN~\cite{wang2021beta}, PyRAT~\cite{girardsatabin2022caisar}, 
to name but a few. Unlike the machine learning approaches, it features a formal language for property specification, \vnnl{}. 
Furthermore, an annual 
 competition VNN-COMP develops common standards for this domain~\cite{BrixMBJL23,abs-2312-16760}. 

However, there are several fundamental problems that prevent these emerging ideas from developing to full fruition. 
Firstly, both the machine learning and verification communities assume that \emph{in theory} a neural network can be optimised for the desirable verification property. 
However, without any programming language support to ensure this formally, discrepancies between machine learning objectives
and verification objectives have been found in the literature, even for simple robustness properties~\cite{CKDKKAE22}. In Fig.~\ref{fig:CV}, this problem is depicted by distinguishing the two versions of \textbf{NN Property} and \textbf{NN Property$^*$} and a dotted line between them.   The desirable solution is to have a single language with the relevant specification, which is then compiled down to either verification or machine learning backends.

Similarly, discrepancies have been reported
 between different representations of neural networks~\cite{Jia2021floating}, e.g., using real numbers in verification and floating point numbers in training. In Fig.~\ref{fig:CV}, this problem is depicted by showing two potentially disagreeing implementations, \textbf{Implementation} and \textbf{Implementation$^*$}. 
 Ideally, we should be able to 
verify the actual programs, and not their idealised descriptions. Or, as an equally acceptable alternative, the solid arrow between two implementations in Fig.~\ref{fig:CV} should be reversed in the other direction -- 
 ensure that the guarantees concerning the verified neural networks extend to their actual implementations, thus establishing the connection along the bottom dotted arrow in Fig.~\ref{fig:CV}.

Finally, neural networks are rarely implemented as stand-alone programs. More often, they
are embedded into larger system development that, in turn, may have its own specification and verification regimes. 
Although the idea of a verified neural network controller is not itself new to the cyber-physical system research (cf. \S~\ref{sec:other}), the programming language support for verification of such systems is a nascent field~\cite{DBLP:journals/corr/abs-2402-10998,DBLP:journals/corr/abs-2405-14058}. 

In this light, we believe it is time to discuss how the verification and synthesis of safe neural networks fit together with general programming practices.
In this ``Fresh Perspectives'' paper,  we give an overview of the current state of the art in implementing neural network verification and explain the challenges the neural network verification community currently faces (Sec.~\ref{sec:soa}). We do so by tracing different parts of the diagram in Fig.~\ref{fig:CV}, and explaining the nature of the discrepancies in its different parts, from the programming language point of view. We wrap up this paper by suggesting possible ways the programming language community can help improve the state of the art (Sec.~\ref{sec:future}).

\vspace{0.5em}

\vspace{-1em}
\section{Neural Network Verification Properties}
\label{sec:benchmarks}

The problem of defining verification properties for neural networks has received substantial attention. Verification approaches started with neural networks deployed as controllers in autonomous systems~\cite{PT10,katz2017reluplex}.  With time, they were generalised to cover data-dependent verification properties such as robustness~\cite{huang2017safety,ehlers2017atva,dutta2018nfm,xiang2018output}. A set of standard benchmarks is revised and updated annually at the VNN-COMP; the competition reports~\cite{brix2023first,brix2024fifthinternationalverificationneural} provide a thorough overview of them. 
 
Neural network verification properties can be divided into three categories.
\begin{enumerate}
    \item \textbf{Geometric properties.} These properties are based on the geometry of the data manifold without any appeal to its possible semantic meaning. One such property is (local) \textit{robustness}, whose definition is given in Equation \ref{robustness} (see also the additional examples in~\cite{casadio2022robust}). Another related property is (local) \textit{equivalence}~\cite{Teuber2021equivalence}, which constrains the output of two different networks to be similar under the same input, either in absolute value ($\epsilon$-equivalence) or class prediction (top-$k$ equivalence).
    \item \textbf{Hyper-properties.} These properties require guarantees for any input, rather than just those close to the data manifold. Classic examples are global robustness~\cite{Prach_2024_CVPR} and global equivalence~\cite{lohar2023milp}. A more recent example of such properties is \textit{confidence-based robustness}~\cite{athavale2024}, which allows for some non-robust behaviour, but only for inputs close to the decision boundary. The latter complicates the specification and verification process in interesting ways (see Sec. \ref{sec:future}).
    \item \textbf{Domain-specific properties.} These properties are based on the presumed semantics of the data on which the neural network is trained.   Usually, they take the form of admissible intervals on the input and output vector values.
\end{enumerate}

The ACAS Xu challenge (the oldest neural network verification benchmark) best illustrates this third class of properties. It takes a neural network that models an aircraft controller: based on five input measurements between the own ship and an intruder (distance, angles, relative speeds), the neural network outputs one of five advisory actions (strong/weak left or right, clear of conflict).  

When the benchmark was introduced in~\cite{katz2017reluplex}, nine properties were formulated by the engineers who designed the collision avoidance software. For instance, Property 3 states that \emph{if the intruder is directly ahead and is moving towards the own ship, the network will not advise clear of conflict.} When written in the \vnnl{} query language~\cite{demarchi2023supporting} (see Sec. \ref{sec:spec}), the property is translated to real-valued intervals on the five input measurements and a constraint on the output prediction.

\section{Neural Network Verification: State of the Art}\label{sec:soa}

In this section, we describe the state of the art in neural network verification, from the perspective of the existing programming language support, rather than the existing verification algorithms. For the latter, the tutorial~\cite{albarghouthi2021introductionneuralnetworkverification} is available.
We will proceed by tracing different arrows of Figure~\ref{fig:CV} and explaining the existing discrepancies and solutions.

\subsection{\label{sec:spec} Verification pipeline}
\vspace{0.5em}

\begin{figure}[t!]
\includegraphics[width=\textwidth]{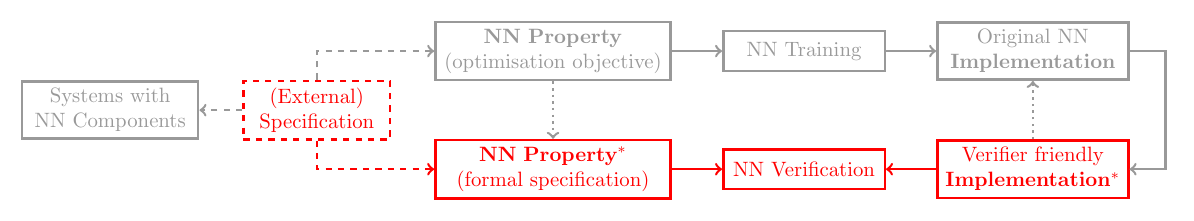}
\caption{\footnotesize{\emph{Schematic representation of the neural network verification pipeline.}}}\label{fig:verif}
\end{figure}

\subsubsection{Neural Network Verification Problem.} Let us start with describing the common verification pipeline illustrated in Fig.~\ref{fig:verif}. Given a trained neural network $f :\R^m \rightarrow \R^n$ and some network property $\networkProperty$, the \emph{Neural Network Verification Problem} is the problem of deciding whether $\networkProperty(f)$ holds. Current verifiers assume using a special format ---  \onnx (standing for \emph{Open Neural Network Exchange})~\cite{onnx2022} --- to represent the neural networks. Thus, in reality, we verify $\networkProperty(f^*)$, where $f^*$ is obtained from $f$ by \onnx translation. 

The verifiers typically consider properties defining a precondition on the network inputs and a postcondition on its outputs.
Both conditions are most commonly linear (e.g., defined using linear bounds) and represent safe regions. Formally, let  $\networkProperty \coloneq \langle P, Q \rangle$ where $P:\mathbb{R}^m\rightarrow\{\top,\bot\}$ and $Q:\mathbb{R}^n\rightarrow\{\top,\bot\}$.
The neural network verification problem is then deciding whether $\forall x\in\mathbb{R}^m: P(x)\Rightarrow Q(f^*(x))$. Neural network verification algorithms then attempt to find a counterexample (i.e., $x\in\mathbb{R}^m$ such that $P(x)\wedge \lnot Q(f^*(x))$) or conclude there is none.
Several \emph{neural network verifiers} are currently available to solve such verification problems: e.g. Marabou~\cite{katz2019marabou,Wu24}, $\alpha\beta$-CROWN~\cite{wang2021beta}, PyRAT~\cite{girardsatabin2022caisar}, NNV~\cite{tran2020cav_tool,lopez2023nnv} and ERAN~\cite{singh2019abstract}. 
Since 2020, an annual International Verification of Neural Networks Competition (VNN-COMP) has been held, and has played an important role in consolidating the new research community and developing standards for this domain~\cite{BrixMBJL23,abs-2312-16760}.

\subsubsection{Mainstream specification languages.}
\label{sec:query-language}
Most neural network verifiers have a basic query language for representing individual queries. 
These formats are invariably simple enough so that the type-system is implicit rather than explicit and they possess no capability to abstract over definitions. 
The \emph{de-facto} standard is the \vnnl{} query language~\cite{demarchi2023supporting}
which is used in \vnnc~\cite{BaLiJo21}. 
The language is a subset of the QFLRA fragment of the \smtl{} language, an S-expression based language widely used in the SMT verification community as a standard input for SMT provers~\cite{barrett2016satisfiability}. The goal of \vnnl{} is to
model first-order logic properties on the inputs and outputs of neural networks. 
Fig.~\ref{fig:spec} illustrates a snippet of robustness specification written in \vnnl{}. As can be seen, \vnnl specification itself does not explicitly talk about the functions $f$ or $f^*$, rather it is assuming that the property will be used to verify the function $f^*$ provided in a separate ONNX file. Thus,
\vnnl{} and \onnx together serve as a specification for $\networkProperty(f^*)$.

\begin{figure}[t!]
\centering
\begin{subfigure}{0.38\textwidth}
\begin{Verbatim}[commandchars=\\\{\}]
\PYG{p}{(}\PYG{n}{declare}\PYG{o}{\PYGZhy{}}\PYG{n}{const}\PYG{+w}{ }\PYG{k+kt}{X\PYGZus{}0}\PYG{+w}{ }\PYG{k+kt}{Real}\PYG{p}{)}
\PYG{p}{(}\PYG{n}{declare}\PYG{o}{\PYGZhy{}}\PYG{n}{const}\PYG{+w}{ }\PYG{k+kt}{X\PYGZus{}1}\PYG{+w}{ }\PYG{k+kt}{Real}\PYG{p}{)}
\PYG{o}{...}
\PYG{p}{(}\PYG{n}{declare}\PYG{o}{\PYGZhy{}}\PYG{n}{const}\PYG{+w}{ }\PYG{k+kt}{X\PYGZus{}791}\PYG{+w}{ }\PYG{k+kt}{Real}\PYG{p}{)}
\PYG{p}{(}\PYG{n}{declare}\PYG{o}{\PYGZhy{}}\PYG{n}{const}\PYG{+w}{ }\PYG{k+kt}{Y\PYGZus{}0}\PYG{+w}{ }\PYG{k+kt}{Real}\PYG{p}{)}
\PYG{o}{...}
\PYG{p}{(}\PYG{n}{declare}\PYG{o}{\PYGZhy{}}\PYG{n}{const}\PYG{+w}{ }\PYG{k+kt}{Y\PYGZus{}8}\PYG{+w}{ }\PYG{k+kt}{Real}\PYG{p}{)}
\PYG{p}{(}\PYG{n}{declare}\PYG{o}{\PYGZhy{}}\PYG{n}{const}\PYG{+w}{ }\PYG{k+kt}{Y\PYGZus{}9}\PYG{+w}{ }\PYG{k+kt}{Real}\PYG{p}{)}
\end{Verbatim}
\end{subfigure}%
~
\begin{subfigure}{0.4\textwidth}
\begin{Verbatim}[commandchars=\\\{\}]
\PYG{p}{(}\PYG{n}{assert}\PYG{+w}{ }\PYG{p}{(}\PYG{o}{\PYGZlt{}=}\PYG{+w}{ }\PYG{k+kt}{X\PYGZus{}0}\PYG{+w}{ }\PYG{l+m+mf}{0.0}\PYG{p}{))}
\PYG{p}{(}\PYG{n}{assert}\PYG{+w}{ }\PYG{p}{(}\PYG{o}{\PYGZgt{}=}\PYG{+w}{ }\PYG{k+kt}{X\PYGZus{}0}\PYG{+w}{ }\PYG{l+m+mf}{0.0}\PYG{p}{))}
\PYG{o}{...}
\PYG{p}{(}\PYG{n}{assert}\PYG{+w}{ }\PYG{p}{(}\PYG{o}{\PYGZlt{}=}\PYG{+w}{ }\PYG{k+kt}{X\PYGZus{}791}\PYG{+w}{ }\PYG{o}{\PYGZhy{}}\PYG{l+m+mf}{58.231295852661134}\PYG{p}{))}
\PYG{p}{(}\PYG{n}{assert}\PYG{+w}{ }\PYG{p}{(}\PYG{o}{\PYGZgt{}=}\PYG{+w}{ }\PYG{k+kt}{X\PYGZus{}791}\PYG{+w}{ }\PYG{o}{\PYGZhy{}}\PYG{l+m+mf}{75.58388969421387}\PYG{p}{))}
\PYG{o}{...}
\PYG{p}{(}\PYG{n}{assert}\PYG{+w}{ }\PYG{p}{(}\PYG{o}{\PYGZgt{}=}\PYG{+w}{ }\PYG{k+kt}{Y\PYGZus{}5}\PYG{+w}{ }\PYG{k+kt}{Y\PYGZus{}1}\PYG{p}{))}
\PYG{p}{(}\PYG{n}{assert}\PYG{+w}{ }\PYG{p}{(}\PYG{o}{\PYGZgt{}=}\PYG{+w}{ }\PYG{k+kt}{Y\PYGZus{}5}\PYG{+w}{ }\PYG{k+kt}{Y\PYGZus{}3}\PYG{p}{))}
\end{Verbatim}
\end{subfigure}    
\caption{\footnotesize{\emph{Snippet of robustness specification in VNN-Lib for an image data set that has input of dimension $792$ and $10$ classes. The specification assumes an external definition of $f^*: \R^{792} \to \R^{10}$.}}}\label{fig:spec}
\end{figure}

%
%


From a programming language perspective, there are several issues with the \vnnl{} format as a language for expressing specifications.
\begin{enumerate}

\item \textbf{Lack of expressivity.} \vnnl and \onnx are simply not expressive enough to represent all the specifications users want to write. 
For example, the \vnnl and \onnx  formats can only refer to a single neural network at a time, which makes encoding specifications where one needs to express
properties on several neural networks at once impossible. Similarly, hyperproperties~\cite{athavale2024,christakis2022} cannot be specified in \vnnl without special tooling, and neither can properties involving hidden neurons. Finally, \vnnl only supports satisfaction queries, meaning the specification writer must manually negate universal queries before being encoded.
\item \textbf{Lack of conciseness.} The lack of abstraction and the limitation that variables cannot represent multi-dimensional tensors means that more complex properties cannot be represented concisely. 
Consequently, the length of the queries tend to scale with the dimensions of inputs and outputs of the network, even when the property can be expressed concisely in mathematics in constant space. For example, the full specification in Fig.~\ref{fig:spec} that encodes the single line of Eq.~\ref{robustness} is a couple of thousand lines long.
\item \textbf{Lack of rigour.}  \vnnl does not have a formally defined semantics, nor does it even formally define its own syntax. 
Consequently, it is difficult for users to check whether their specification in \vnnl is correct or compliant, and impossible to prove the soundness of tools that either consume or generate \vnnl.
Furthermore, the \onnx{} format that \vnnl{} relies on, also lacks a formal semantics. For example, the \onnx documentation for the convolution operator\footnote{\url{https://onnx.ai/onnx/operators/onnx__Conv.html}, accessed \emph{21-09-2024}} has no proper mathematical specification for the semantics of the operator, describing it only with the single sentence ``The convolution operator consumes an input tensor and a filter, and computes the output''.
Other ONNX operator descriptions like those of Convolution, Maxpool, or Add (for broadcasting) refer to external sources like Numpy, PyTorch or Tensorflow for more implementation details.

\item \textbf{Lack of dynamic bindings to datasets.} Crucial to most attempts to specify ``correctness'' of a neural network is the notion of the \emph{data manifold}, i.e., the distribution of inputs that the neural network will actually encounter during operation. 
Usually, the data manifold is only a small subset of the actual input space. By definition, the network should never encounter inputs that lie off the data-manifold during normal operation. If it does, there is no reason to require any particular behaviour from the network, and consequently, specifications should only quantify over inputs that lie on the manifold.
The problem is that, in most cases, there is no precise mathematical definition of the data manifold. 
Therefore, the most common approach is for the specification to approximate the manifold as the union of ``small'' regions around each input in the training dataset. Unfortunately, the training datasets themselves are frequently huge,
anywhere from thousands to hundreds of millions of items. Therefore, it is infeasible to directly express the dataset in the specification. 

\end{enumerate}

\noindent This lack of rigour of the underlying specification format has been recognised as a major problem. A recent effort in the ONNX community has led to the creation of a ONNX Safety-Related Profile working group\footnote{\url{https://github.com/ericjenn/working-groups/blob/ericjenn-srpwg-wg1/safety-related-profile/README.md}} which aims to elaborate a dedicated ONNX profile for safety-related systems. While still embryonic, this working group might answer some of the issues highlighted above. 

To work around the remaining problems, the natural solution is to allow users to represent their specifications in a higher-level specification language, connecting the neural network specification to the language of the larger system in which it is embedded. 
Moreover, the specification language must provide some mechanism for dynamically binding variables to existing datasets in standard formats used by machine learning practitioners.



\subsection{Prototypes of New Specification Languages}\label{sec:VC}

\begin{figure}[t!]
\begin{subfigure}{0.58\textwidth}

\begin{Verbatim}[commandchars=\\\{\},fontsize=\scriptsize]
\PYG{n}{theory} \PYG{n+nc}{MNIST}

  \PYG{n}{use} \PYG{n}{ieee\PYGZus{}float}\PYG{o}{.}\PYG{n+nc}{Float64}
  \PYG{n}{use} \PYG{n}{caisar}\PYG{o}{.}\PYG{n}{types}\PYG{o}{.}\PYG{n+nc}{Float64WithBounds} \PYG{k}{as} \PYG{n+nc}{Feature}
  \PYG{n}{use} \PYG{n}{caisar}\PYG{o}{.}\PYG{n}{types}\PYG{o}{.}\PYG{n+nc}{IntWithBounds} \PYG{k}{as} \PYG{n+nc}{Label}
  \PYG{n}{use} \PYG{n}{caisar}\PYG{o}{.}\PYG{n}{model}\PYG{o}{.}\PYG{n+nc}{Model}

  \PYG{n}{use} \PYG{n}{caisar}\PYG{o}{.}\PYG{n}{dataset}\PYG{o}{.}\PYG{n+nc}{CSV}
  \PYG{n}{use} \PYG{n}{caisar}\PYG{o}{.}\PYG{n}{robust}\PYG{o}{.}\PYG{n+nc}{ClassRobustCSV}

  \PYG{n}{constant} \PYG{n}{model\PYGZus{}filename}\PYG{o}{:} \PYG{k+kt}{string}
  \PYG{n}{constant} \PYG{n}{dataset\PYGZus{}filename}\PYG{o}{:} \PYG{k+kt}{string}

  \PYG{n}{constant} \PYG{n}{label\PYGZus{}bounds}\PYG{o}{:} \PYG{n+nn}{Label}\PYG{p}{.}\PYG{n}{bounds} \PYG{o}{=}
    \PYG{n+nn}{Label}\PYG{p}{.}\PYG{o}{\PYGZob{}} \PYG{n}{lower} \PYG{o}{=} \PYG{l+m+mi}{0}\PYG{o}{;} \PYG{n}{upper} \PYG{o}{=} \PYG{l+m+mi}{9} \PYG{o}{\PYGZcb{}}

  \PYG{n}{constant} \PYG{n}{feature\PYGZus{}bounds}\PYG{o}{:} \PYG{n+nn}{Feature}\PYG{p}{.}\PYG{n}{bounds} \PYG{o}{=}
    \PYG{n+nn}{Feature}\PYG{p}{.}\PYG{o}{\PYGZob{}} \PYG{n}{lower} \PYG{o}{=} \PYG{o}{(}\PYG{l+m+mi}{0}\PYG{o}{.}\PYG{l+m+mi}{0}\PYG{o}{:}\PYG{n}{t}\PYG{o}{);} \PYG{n}{upper} \PYG{o}{=} \PYG{o}{(}\PYG{l+m+mi}{1}\PYG{o}{.}\PYG{l+m+mi}{0}\PYG{o}{:}\PYG{n}{t}\PYG{o}{)} \PYG{o}{\PYGZcb{}}

\PYG{o}{[...]}
  \PYG{n}{predicate} \PYG{n}{robust} \PYG{o}{(}\PYG{n}{f\PYGZus{}bounds}\PYG{o}{:} \PYG{n+nn}{Feature}\PYG{p}{.}\PYG{n}{bounds}\PYG{o}{)}
                   \PYG{o}{(}\PYG{n}{l\PYGZus{}bounds}\PYG{o}{:} \PYG{n+nn}{Label}\PYG{p}{.}\PYG{n}{bounds}\PYG{o}{)}
                   \PYG{o}{(}\PYG{n}{m}\PYG{o}{:} \PYG{n}{model}\PYG{o}{)} \PYG{o}{(}\PYG{n}{eps}\PYG{o}{:} \PYG{n}{t}\PYG{o}{)}
                   \PYG{o}{(}\PYG{n}{l}\PYG{o}{:} \PYG{n+nn}{Label}\PYG{p}{.}\PYG{n}{t}\PYG{o}{)}
                   \PYG{o}{(}\PYG{n}{e}\PYG{o}{:} \PYG{n+nn}{FeatureVector}\PYG{p}{.}\PYG{n}{t}\PYG{o}{)} \PYG{o}{=}
    \PYG{n}{forall} \PYG{n}{perturbed\PYGZus{}e}\PYG{o}{:} \PYG{n+nn}{FeatureVector}\PYG{p}{.}\PYG{n}{t}\PYG{o}{.}
      \PYG{n}{has\PYGZus{}length} \PYG{n}{perturbed\PYGZus{}e} \PYG{o}{(}\PYG{n}{length} \PYG{n}{e}\PYG{o}{)} \PYG{o}{\PYGZhy{}\PYGZgt{}}
      \PYG{n+nn}{FeatureVector}\PYG{p}{.}\PYG{n}{valid} \PYG{n}{f\PYGZus{}bounds} \PYG{n}{perturbed\PYGZus{}e} \PYG{o}{\PYGZhy{}\PYGZgt{}}
      \PYG{k}{let} \PYG{n}{perturbation} \PYG{o}{=} \PYG{n}{perturbed\PYGZus{}e} \PYG{o}{\PYGZhy{}} \PYG{n}{e} \PYG{k}{in}
      \PYG{n}{bounded\PYGZus{}by\PYGZus{}epsilon} \PYG{n}{perturbation} \PYG{n}{eps} \PYG{o}{\PYGZhy{}\PYGZgt{}}
      \PYG{n}{advises} \PYG{n}{l\PYGZus{}bounds} \PYG{n}{m} \PYG{n}{perturbed\PYGZus{}e} \PYG{n}{l}

\PYG{o}{[...]}
  \PYG{n}{goal} \PYG{n}{robustness}\PYG{o}{:}
    \PYG{k}{let} \PYG{n}{nn} \PYG{o}{=} \PYG{n}{read\PYGZus{}model} \PYG{n}{model\PYGZus{}filename} \PYG{k}{in}
    \PYG{k}{let} \PYG{n}{dataset} \PYG{o}{=} \PYG{n}{read\PYGZus{}dataset} \PYG{n}{dataset\PYGZus{}filename} \PYG{k}{in}
    \PYG{k}{let} \PYG{n}{eps} \PYG{o}{=} \PYG{o}{(}\PYG{l+m+mi}{0}\PYG{o}{.}\PYG{l+m+mi}{010000000}\PYG{o}{...:}\PYG{n}{t}\PYG{o}{)} \PYG{k}{in}
    \PYG{n}{robust} \PYG{n}{feature\PYGZus{}bounds} \PYG{n}{label\PYGZus{}bounds} \PYG{n}{nn} \PYG{n}{dataset} \PYG{n}{eps}

\PYG{k}{end}
\end{Verbatim}
\caption{CAISAR}
\end{subfigure}%
~ 
\begin{subfigure}{0.38\textwidth}
\begin{Verbatim}[commandchars=\\\{\},fontsize=\scriptsize]
\PYG{k}{type} \PYG{n+nc}{Label} \PYG{o}{=} \PYG{n+nc}{Index} \PYG{l+m+mi}{10}
\PYG{k}{type} \PYG{n+nc}{Image} \PYG{o}{=} \PYG{n+nc}{Tensor} \PYG{n+nc}{Rat} \PYG{o}{[}\PYG{l+m+mi}{28}\PYG{o}{,} \PYG{l+m+mi}{28}\PYG{o}{]}

\PYG{o}{@}\PYG{n}{network}
\PYG{n}{mnist} \PYG{o}{:} \PYG{n+nc}{Image} \PYG{o}{\PYGZhy{}\PYGZgt{}} \PYG{n+nc}{Tensor} \PYG{n+nc}{Rat} \PYG{o}{[}\PYG{l+m+mi}{10}\PYG{o}{]}

\PYG{n}{validImage} \PYG{o}{:} \PYG{n+nc}{Image} \PYG{o}{\PYGZhy{}\PYGZgt{}} \PYG{n+nc}{Bool}
\PYG{n}{validImage} \PYG{n}{x} \PYG{o}{=} \PYG{n}{forall} \PYG{n}{i} \PYG{n}{j} \PYG{o}{.}
  \PYG{l+m+mi}{0} \PYG{o}{\PYGZlt{}=} \PYG{n}{x} \PYG{o}{!} \PYG{n}{i} \PYG{o}{!} \PYG{n}{j} \PYG{o}{\PYGZlt{}=} \PYG{l+m+mi}{1}

\PYG{n}{advises} \PYG{o}{:} \PYG{n+nc}{Image} \PYG{o}{\PYGZhy{}\PYGZgt{}} \PYG{n+nc}{Label} \PYG{o}{\PYGZhy{}\PYGZgt{}} \PYG{n+nc}{Bool}
\PYG{n}{advises} \PYG{n}{x} \PYG{n}{i} \PYG{o}{=} \PYG{n}{forall} \PYG{n}{j} \PYG{o}{.}
  \PYG{n}{j} \PYG{o}{!=} \PYG{n}{i} \PYG{o}{=\PYGZgt{}} \PYG{n}{mnist} \PYG{n}{x} \PYG{o}{!} \PYG{n}{i} \PYG{o}{\PYGZgt{}} \PYG{n}{mnist} \PYG{n}{x} \PYG{o}{!} \PYG{n}{j}

\PYG{o}{@}\PYG{n}{parameter}
\PYG{n}{epsilon} \PYG{o}{:} \PYG{n+nc}{Rat}

\PYG{n}{boundedByEpsilon} \PYG{o}{:} \PYG{n+nc}{Image} \PYG{o}{\PYGZhy{}\PYGZgt{}} \PYG{n+nc}{Bool}
\PYG{n}{boundedByEpsilon} \PYG{n}{x} \PYG{o}{=} \PYG{n}{forall} \PYG{n}{i} \PYG{n}{j} \PYG{o}{.}
  \PYG{o}{\PYGZhy{}}\PYG{n}{epsilon} \PYG{o}{\PYGZlt{}=} \PYG{n}{x} \PYG{o}{!} \PYG{n}{i} \PYG{o}{!} \PYG{n}{j} \PYG{o}{\PYGZlt{}=} \PYG{n}{epsilon}

\PYG{n}{robust} \PYG{o}{:} \PYG{n+nc}{Label} \PYG{o}{\PYGZhy{}\PYGZgt{}} \PYG{n+nc}{Image} \PYG{o}{\PYGZhy{}\PYGZgt{}} \PYG{n+nc}{Bool}
\PYG{n}{robust} \PYG{n}{label} \PYG{n}{image} \PYG{o}{=} \PYG{n}{forall} \PYG{n}{perturbation} \PYG{o}{.}
  \PYG{n}{boundedByEpsilon} \PYG{n}{perturbation} \PYG{o+ow}{and}
  \PYG{n}{validImage} \PYG{o}{(}\PYG{n}{perturbation} \PYG{o}{+} \PYG{n}{image}\PYG{o}{)} \PYG{o}{=\PYGZgt{}}
  \PYG{n}{advises} \PYG{n}{label} \PYG{o}{(}\PYG{n}{perturbation} \PYG{o}{+} \PYG{n}{image}\PYG{o}{)}

\PYG{o}{@}\PYG{n}{parameter}\PYG{o}{(}\PYG{n}{infer}\PYG{o}{=}\PYG{n+nc}{True}\PYG{o}{)}
\PYG{n}{n} \PYG{o}{:} \PYG{n+nc}{Nat}

\PYG{o}{@}\PYG{n}{dataset}
\PYG{n}{images} \PYG{o}{:} \PYG{n+nc}{Tensor} \PYG{n+nc}{Image} \PYG{o}{[}\PYG{n}{n}\PYG{o}{]}

\PYG{o}{@}\PYG{n}{dataset}
\PYG{n}{labels} \PYG{o}{:} \PYG{n+nc}{Tensor} \PYG{n+nc}{Label} \PYG{o}{[}\PYG{n}{n}\PYG{o}{]}

\PYG{o}{@}\PYG{n}{property}
\PYG{n}{robustness} \PYG{o}{:} \PYG{n+nc}{Tensor} \PYG{n+nc}{Bool} \PYG{o}{[}\PYG{n}{n}\PYG{o}{]}
\PYG{n}{robustness} \PYG{o}{=} \PYG{n}{foreach} \PYG{n}{i} \PYG{o}{.}
  \PYG{n}{robustAround} \PYG{o}{(}\PYG{n}{images} \PYG{o}{!} \PYG{n}{i}\PYG{o}{)} \PYG{o}{(}\PYG{n}{labels} \PYG{o}{!} \PYG{n}{i}\PYG{o}{)}
\end{Verbatim}
\caption{Vehicle}
\end{subfigure}    
\caption{\footnotesize{\emph{An extract from a local robustness specification in CAISAR and Vehicle's input languages for the same image dataset described in Fig.~\ref{fig:spec}. Note the ability to reuse predicates and definitions, the conciseness of vector-based operations, and the explicit data set bindings.}}}\label{fig:spec-caisar}
\end{figure}

\noindent In response to the outlined problems, two major attempts have been made to design more principled specification languages for neural network verification.
We outline the essence of both, in turn. 
Fig.~\ref{fig:spec-caisar} provides code snippets for illustration. 

\begin{enumerate} 
\item \textbf{CAISAR.} The \caisar platform~\cite{girardsatabin2022caisar} incorporates a higher-level specification language deriving from WhyML~\cite{filliatre2013why3}. WhyML is a typed first-order language with pattern-matching, polymorphism, and a module system. 
On top of that, \caisar provides additional types of linear algebra structures common in machine learning and compiles the specification back to plain WhyML. Writing a compiler from WhyML to \vnnl is straightforward, allowing \caisar to target all state-of-the-art solvers from one single specification. It can also deal with specifications involving multiple neural networks and dynamically bind variables to concrete datasets. However, it can be argued that the composability of WhyML is limited, and the lack of dependent types prevents the modelling of important properties (for instance, encoding the dimension of inputs directly in their types could prevent common runtime errors). 

\item \textbf{Vehicle.}
The Vehicle specification language~\cite{DaggittKKA0SCCL23,DBLP:journals/corr/abs-2401-06379,Vehicle} is a higher-order and dependently-typed functional language. The language aims to be able to express a full range of specifications and to that end it contains quantifiers as first-class language constructs, conditionals and higher-order functions over tensors such as maps and folds.
The language's dependently typed nature allows the user to encode richer properties and includes tensor size constraints that can be checked before verification by the type-checker. Vehicle also has a backend that allows connecting proofs of neural network properties to larger system specifications in Agda~\cite{DBLP:journals/corr/abs-2401-06379}. 
However, unlike \caisar{}, it connects to far fewer tools and cannot allow multiple solvers to work together.
\end{enumerate}

These two languages solve the problems outlined in Sec.~\ref{sec:query-language} and provide a concrete implementation. Note, in particular, that both manage data set bindings, neural network bindings, and data validity checks in clear, explicit ways. By doing this, they are essentially building the specification languages on top of the existing pipelines: in Figs.~\ref{fig:CV} and \ref{fig:verif}, this is depicted by a dashed ``Specification'' box towards the left side.  Other specification languages exist, like NeSAL~\cite{xie2022neuro} (which has no implementation) or DNNP~\cite{shriver2021dnnv} (lacking quantifiers and strong typing).





\vspace{-1em}
\subsection{The Embedding Gap\label{sec:embgap}}

\begin{figure}[t!]
\includegraphics[width=\textwidth]{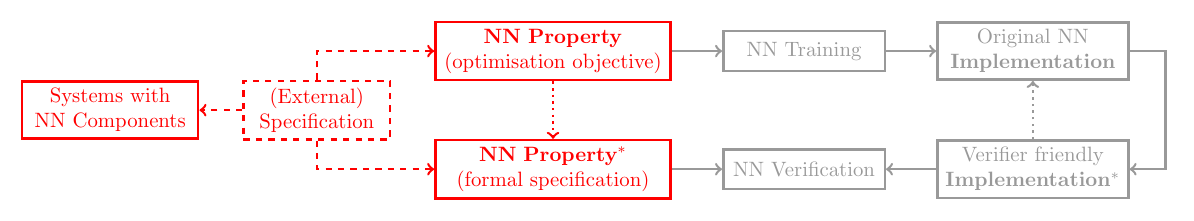}
\caption{\footnotesize{\emph{Schematic representation of the embedding gap.}}}\label{fig:embed}
\end{figure}

%
We now consider the influence of larger system verification on the neural network verification pipeline (see Fig.~\ref{fig:embed}). Consider a purely symbolic program $\system(\cdot)$, whose completion requires computing a complex, unknown function~$\hypothesis : \problemSpace \rightarrow \resultSpace$ that maps objects in the \emph{problem input space} $\problemSpace$ to those in the \emph{problem output space}~$\resultSpace$. Given an embedding function $e : \problemSpace \rightarrow \R^m$ and an unembedding function $u : \R^n \rightarrow \resultSpace$, we can approximate~$\hypothesis$ by training a neural network~${f : \R^m \rightarrow \R^n}$ such that $u \circ f \circ e \approx \hypothesis$. We refer to $u \circ f \circ e$ as the \emph{solution}, and refer to $\R^m$ and $\R^n$ as the \emph{embedding input space} and \emph{embedding output space} respectively. Unlike objects in the problem space, the vectors in the embedding space are often not directly interpretable. The complete program is then modelled as $\system(u \circ f \circ e)$. 
Examples of $u$ and $e$ would be the normalization of inputs, resizing operations for images, or data augmentation operations that are commonplace in machine learning pipelines.

Our end goal is to prove that~$\system(u \circ f \circ e)$ satisfies a property~$\systemProperty$, which we will refer to as the \emph{program property}. The natural way to proceed is to establish a \emph{solution property}~$\solutionProperty$ and a \emph{network property}~$\networkProperty$ such that the proof of~$\systemProperty$ is decomposable into the following three lemmas:
\begin{align}
    \label{eq:networkProperty}
    \networkProperty(f)
    \qquad \qquad \qquad \qquad \qquad \qquad \qquad \qquad \qquad \qquad \\
    \label{eq:solutionProperty}
    \forall g : \networkProperty(g) \Rightarrow \solutionProperty(u \circ g \circ e)
    \qquad \qquad \qquad \qquad \quad                                     \\ \qquad \qquad \qquad \qquad
    \label{eq:systemProperty}
    \forall h : \solutionProperty(h) \Rightarrow
    \systemProperty(\system(h))
\end{align}
i.e. Lemma~\ref{eq:networkProperty} proves that the network~$f$ obeys the network property~$\networkProperty$, then Lemma~\ref{eq:solutionProperty} proves that this implies $u \circ f \circ e$, the neural network lifted to the problem space, obeys the solution property~$\solutionProperty$, and finally Lemma~\ref{eq:systemProperty} proves that this implies $\system(u \circ f \circ e)$, the neuro-symbolic program, obeys the program property~$\systemProperty$.

The first issue that we run into is what we call the \emph{embedding gap}.
In $\systemProperty$, users would like to be able to model data that potentially has non-trivial semantics (for example, featuring both continuous and discrete parameters of a cyber-physical system such as velocity, stopping distance, switches etc.). However, in $\networkProperty$, all values must be represented as continuous real vectors (in actuality, at the training phase, floating-point vectors, cf. Sec.~\ref{sec:integrity}). A function from the latter to the former must be highly non-surjective.

For example, consider an input type with two values, ‘Yes’ and ‘No’, encoded as real values ‘0.0’ and ‘1.0’ correspondingly.
In the low-level query, one can encode that this input variable can only take two possible values using a disjunctive constraint ($x = 0.0 \vee x = 1.0$), but this does not scale well as the number of constructors in the data type grows, as each disjunction drastically increases the cost of verification.
Instead, the most common current solution is to encode this as a single non-disjunctive constraint, $0.0 \leq x \leq 1$. In this case, the problem is that floating-point numbers may contain other values (e.g., `0.005', `0.97'), which are meaningless in the chosen domain.

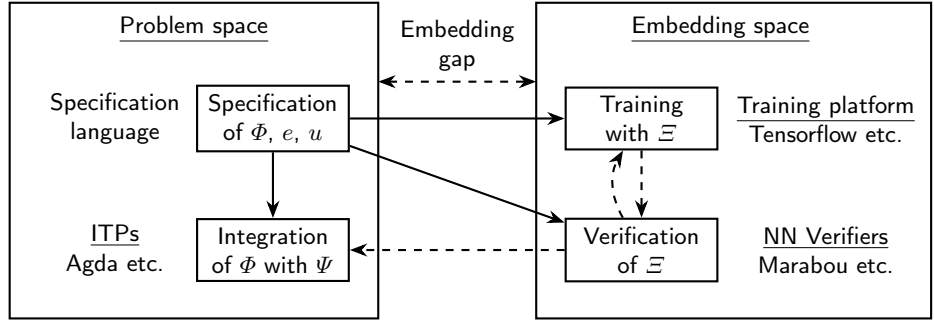
\begin{figure}[t] 
  	\vspace{-1em}
\begin{tikzpicture}[thick, scale=0.7, font=\sffamily\footnotesize,
    set/.style = {circle,
        maximum size = 2cm}, -> /.tip = Stealth]

\newcommand{\rowOneY}{11.8};
\newcommand{\rowTwoY}{9.3};
\newcommand{\colFourX}{15.5};

\draw[draw=black]  (7,14) rectangle ++ (-7,-6);

\draw[draw=black]  (17.5,14) rectangle ++ (-7.5,-6);

\path [<->,dashed] (7, 12.5) edge (10, 12.5);

\node[align=center, fill=white, inner sep=0, outer sep=0] at (8.5,13.2) {Embedding \\ gap};

\node[align=center] at (3.5,13.5) {\underline{Problem space}};

\node[align=center] at (13.5,13.5) {\underline{Embedding space}};

\node[rectangle, text width=1.8cm, draw, align=center] (A) at (5,\rowOneY) {Specification of $\solutionProperty$, $e$, $u$};
\node[align=center] at (2,\rowOneY) {Specification \\ language};

\node[rectangle, text width=1.8cm, draw, align=center] (B) at (12,\rowOneY) {Training with $\networkProperty$};
\node[align=center] at (\colFourX,\rowOneY) {\underline{Training platform} \\ Tensorflow etc. };

\draw [->] (A) edge (B);

\node[rectangle,draw,text width=1.8cm, align=center] (D) at (12,\rowTwoY) {Verification of $\networkProperty$};
\node[text width=2cm, align=center] at (\colFourX,\rowTwoY) {\underline{NN Verifiers} \\ Marabou etc.};

\draw [->] (A) edge (D);

\node[rectangle,draw,text width=1.8cm, align=center] (E) at (5,\rowTwoY) {Integration of $\solutionProperty$ with $\systemProperty$};
\node[text width=1.7cm, align=center] at (2,\rowTwoY) {\underline{ITPs} \\ Agda etc.};

\draw [->] (A) edge (E);
\draw [->, dashed] (D) edge (E);
\draw [->, dashed] (B) edge (D);

\draw [->, dashed] (D) edge[bend left=30] (B);
\end{tikzpicture}

\caption{\footnotesize{\emph{Outline of Vehicle compiler backends, bridging the Embedding Gap~\cite{DaggittKKA0SCCL23,DBLP:journals/corr/abs-2401-06379}. Dashed lines indicate information flow and solid lines automatic compilation.}}}\label{fig:Vehicle}
\end{figure}

More generally, if users are to express specifications in $\systemProperty$, the high-level specification language must also allow users to specify the embedding and unembedding functions, $e$ and $u$, as part of the specification.
It should then be the responsibility of the compiler to generate suitable low-level queries representing $\networkProperty$.
%
%
However, allowing the user to encode their specifications at the high-level $\solutionProperty$ requires that the specification language compiler must be able to automatically translate from the former to the latter. 
The only existing attempt to provide programming language support for this was made by Vehicle~\cite{DaggittKKA0SCCL23,DBLP:journals/corr/abs-2401-06379,Vehicle} as shown in Fig.~\ref{fig:Vehicle}. In particular, Vehicle proposes a specification language to express $\solutionProperty$, $e$, $u$, and can compile the specification to Agda, in which more general properties of $\system(\cdot)$ can be defined. 

\subsection{The Implementation Gap}
\label{sec:integrity}

\begin{figure}[t!]
\includegraphics[width=\textwidth]{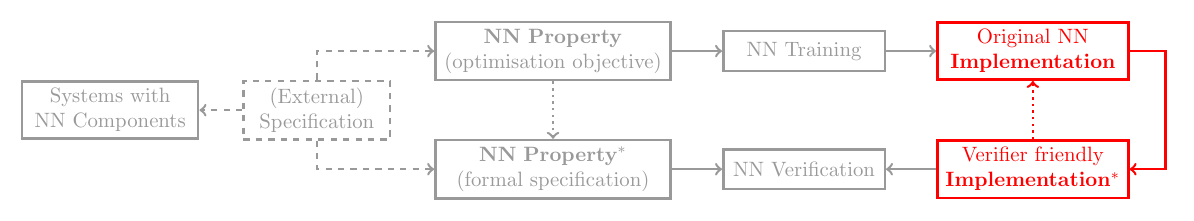}
\caption{\footnotesize{\emph{Schematic representation of the implementation gap.}}}\label{fig:impl}
\end{figure}

In Sec.~\ref{sec:spec} we considered $\networkProperty(f^*)$, where $f^*$ was an  \onnx{} object, possibly obtained by conversion from the original implementation of $f$. 
The \onnx{} format has no backward translation from $f^*$ to $f$, as the diagram in Fig.~\ref{fig:impl} shows. 
However, in the majority of neural network verification publications, authors implicitly assume that obtained verification guarantees about $f^*$  extend to $f$. In this section, we outline a range of problems caused by this and thus trace the right-most section of the diagram illustrated in Figs.~\ref{fig:CV} and \ref{fig:impl}.


\subsubsection{Poor support for neural architecture conversion to \onnx{}.}
\onnx re-implementation of original neural networks remains a largely manual and un-verified procedure, which may be a source of errors. For example,
neural networks contain different types of linear (e.g., fully connected, convolutional) and non-linear (e.g., ReLU, sigmoid, MaxPool) connections. Supporting the formal analysis of a new type of connection typically requires tool developers to add a new dedicated module to the codebase. For example, in verifiers based on abstract interpretation~\cite{singh2019abstract}, this process would involve implementing the abstract transformer for the new type of connection. In SMT-based verification procedures~\cite{katz2017reluplex,wu2022efficient}, the developer would need to implement the encoding, simplification, and satisfiability checking of constraints corresponding to the new connection. This process is tedious, repetitive, and error-prone. For example, the verification code for two-phase activation functions such as ReLU, Leaky ReLU, and absolute values is very similar, yet developers typically need to hard-code separate verification modules for each of these connections.
Ideally, there should be automated conversion procedures with correctness guarantees.  

\vspace{-1em}
\subsubsection{Mismatch in numerical types.}
\label{sec:numerical-types}

Barring experimental architectures that rely on analog computing~\cite{yao2020memristor}, most implementations of neural networks are based on digital platforms that operate with finite-precision types such as integer and floating-point numbers. Effective conversion between real-valued types and finite-precision ones is an active research direction in machine learning~\cite{gholami2022survey}. 

The most ubiquitous 
numerical type in machine learning is the floating-point number~\cite{Li2020Additive,baskin2021nonuniform}. Indeed, the IEEE 754 single precision (32-bit) floating point type~\cite{IEEE754} is the de facto standard of libraries such as Tensorflow\footnote{https://www.tensorflow.org} and Pytorch\footnote{https://pytorch.org/}. 
Efforts to improve over the IEEE 754 standard exist, but they are often relegated to the context of hardware accelerators, where reducing the bit-size of numerical types 
may yield significant gains in terms of speed, memory and power consumption~\cite{NEURIPS2018_335d3d1c,burgess2019floats}.

From the verification perspective, it is well known that the safety certificates produced by real-valued neural network verifiers do not hold for floating-point implementations~\cite{Jia2021floating,zombori2021fooling}. Indeed, Jia and Rinard~\cite{Jia2021floating} propose an algorithm to search for floating-point counterexamples to real-valued safety certificates, thus invalidating them. Similarly, Zombori et al.~\cite{zombori2021fooling} construct neural networks that contain undetectable backdoors, as long as the effects of numerical precision are neglected. Furthermore, the counterexamples produced by real-valued verifiers may not exist on a floating-point implementation of the same neural network, a phenomenon that has been reported on some VNN-COMP benchmarks~\cite{manino2023neurocodebench}.

\vspace{-1em}

\subsubsection{Other sources of non-determinism.}

The current machine learning workflow, from training to inference, is not reproducible across different hardware and software platforms~\cite{pham2020variance,MLSYS2021_5190e987,Zhuang2022randomness,Schloegl2023numerical}. This is due to a variety of reasons:
\begin{enumerate}
    \item \textbf{Non-associativity of floating-point.} It is well-known that floating-point operations are not associative, i.e., $a+(b+c)\neq(a+b)+c$. As such, we can only verify the behaviour of a floating-point neural network if we know the \textit{order} of all its operations. The existing \emph{de-facto} standard ONNX does not include such a level of detail.
    \item \textbf{Parallel execution.} Inference and training of neural networks are often sped up via parallel execution. Whether this is done via SIMD operations, multi-core CPUs, or GPU parallelism, it always introduces non-determinism in the results~\cite{pham2020variance,Schloegl2023numerical}.
    \item \textbf{Auto-selection of primitives.} Modern machine learning compilers like XLA\footnote{https://openxla.org/xla} automatically select the most efficient algorithms depending on the computational load~\cite{pham2020variance}. While PyTorch or Keras present ways to fix the behaviour of the algorithm
    , the ONNX runtime does not. For instance, Sch{\"o}gl et al.~\cite{Schloegl2023numerical} report non-deterministic behaviour in the selection of convolutional algorithms on GPUs, which may alternate between explicit loop, GEMM-based, Winograd and FFT implementations.
    \item \textbf{Runtime optimisations.} Machine learning frameworks may also implement runtime optimisation modifying the structure of the model itself to speed up inference or reduce memory usage, for example by fusing layers together (e.g. convolution and batch normalisation).
    \item \textbf{Non-deterministic training.} The learning process itself is highly non-deterministic. Common sources include: parameter initialisation, data augmentation strategy, batch ordering, and dropout layers~\cite{pham2020variance}.
    \item \textbf{Mathematical library rounding.} A long-standing issue in floating-point computation is incorrect roundings in the standard mathematical library \texttt{math.h}. Technically, the IEEE 754 standard recommends correct roundings~\cite{IEEE754}, and there are efforts to create open-source implementations of \texttt{math.h} that abide by it~\cite{Sibidanov2022coremath}. However, mainstream compilers 
 instead implement a variety of approximately-rounded algorithms~\cite{Brisebarre2024correctly}. \item \textbf{Low-level implementation details.} Furthermore, derived operators such as Softmax may leverage the fact that $\text{softmax}(x + c) = \text{softmax}(x)$ with constant $c$ to increase the precision and avoid overflows. Such details can only be found in the low-level source code,
    even though they severely affect the precision of the computation.
\end{enumerate}
Overall, the end-to-end effects of the above causes of non-determinism cannot be neglected. Indeed, Pham et al.~\cite{pham2020variance} reports a $2.9\%$ difference in accuracy while reproducing the same training run on different platforms. Similarly, Cidon et al.~\cite{MLSYS2021_5190e987} reports a $6\%$ difference in accuracy when considering the whole image recognition pipeline, including camera noise and image processing algorithms.

From the verification perspective, certifying the safety of neural network implementations requires a different approach than high-level neural network verifiers like Marabou~\cite{Wu24} or $\alpha\beta$-CROWN~\cite{wang2021beta}. Indeed, if we had access to the low-level implementation of every library in the machine learning pipeline, we could employ software verifiers~\cite{svcomp2023} for this purpose. Unfortunately, existing software verifiers struggle to cope with the scale and complexity of neural network code~\cite{manino2023neurocodebench,magalhaes2023c2taco,matos2024ceg4n}. In contrast, automated testing approaches are currently more effective~\cite{pmlr-v97-odena19a,Guo2021,Deng2023}, but cannot prove correctness.

\vspace{-0.5em}

\subsubsection{Quantised neural networks.}
Switching to integer types (uniform quantisation)~\cite{gholami2022survey} can help alleviate some of the above problems (e.g. non-associativity of floating point, incorrect rounding) and improve reproducibility.
From the machine learning perspective, a good quantisation scheme maintains the accuracy of the original floating-point neural network.
Usually, 8-bit integers are used, but more aggressive quantisation schemes exist, down to ternary~\cite{He2019ternary} and binary representations~\cite{Qin2020binarised}.
%
 
%
From the verification perspective, integer and binary data types require fundamentally different representations than the real-valued types used by mainstream verifiers such as Marabou~\cite{Wu24} and $\alpha\beta$-CROWN~\cite{wang2021beta}. Existing work on verifying quantised neural networks relies on either the bit-vector SMT theory~\cite{giacobbe2020quantised,Baranowski2020quantised,henzinger2021scalable} or (mixed) integer linear programming (ILP, MILP)~\cite{Mistry2022milp,lohar2023milp,Zhang2023milp,huang2024towards}. In contrast, verifying the robustness of some binarised neural network architectures can be encoded as a satisfiability (SAT) problem~\cite{narodytska2018verifying,Jia2020binarised}. Other binarised architectures can still be encoded as a real-valued verification instance~\cite{amir2021binarised}.

\vspace{-0.5em}
 \subsection{Reliable Proof Production and Checking}\label{sec:proof}
To overcome some of the challenges raised above, neural network verifiers may accompany their results with proof certificates, attesting their soundness using an external and relatively simple checking program (the \emph{proof checker})~\cite{IsBaZhKa22}. Since neural network verifiers are complex software, optimised for performance and speed, their verification is commonly intractable. Thus, proof production replaces the need to directly verify the neural network verifiers, with the need to verify only the proof checker.
When a safety property is violated, neural network verifiers often accompany their results with a counterexample, which can then be checked by its evaluation in the network. However, proving safety (i.e., absence of violation of the property) is not straightforward, as the DNN verification problem is NP-complete even for simple networks~\cite{katz2017reluplex,SaLa21}. Therefore, proving safety is a greater challenge than proving a violation, and thus requires a more complicated proof and, consequently, a more complicated proof checker.  

Proof production mechanism, supporting several piecewise-linear activation functions, was implemented on top of Marabou~\cite{IsBaZhKa22,Wu24}. The proofs produced by Marabou are checked by a proof checker implemented within Marabou. The Marabou proof checker is implemented in C++ and uses floating points arithmetic for its computations.

When using an external proof checker, the reliability of the neural network verifier is dependent on the reliability of the proof checker. Therefore, the proof checker is expected to meet higher standards of reliability, ideally provable soundness. Functional programming languages allow the implementation of a precise checker and formal verification of its soundness. For example, a simply-typed language Imandra was deployed to check proofs produced by Marabou~\cite{desmartin2024certified,DesmartinIPSKK23}.   
This work also shows that computations with precise real arithmetic come at a price of limited performance. This opens up the possibility for a variety of implementations of the same checking algorithm in different programming languages, exploring the trade-off between precision and performance speed.




\subsection{Property-Guided Training}\label{sec:ML}
\vspace{-0.2em}

\begin{figure}[t!]
\includegraphics[width=\textwidth]{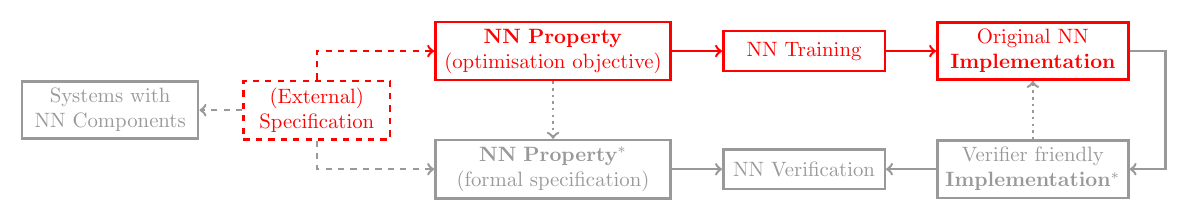}
\caption{\footnotesize{\emph{Schematic representation of the neural network training pipeline.}}}\label{fig:train}
\end{figure}

Finally, we give a brief outline of the state-of-the-art in the property-guided training, which occupies the upper section of Fig.~\ref{fig:train}. This is a booming area in its own right, also known under the umbrella term of \emph{neuro-symbolic AI}.
By pointing out existing programming language discrepancies and solutions, we do not attempt to give a full survey of neuro-symbolic AI, but refer the reader to more comprehensive surveys~\cite{GiunchigliaSL22,hitzler2022neuro}.

In the introduction, we have already outlined the evolution from adversarial training (seen as training for the robustness property specifically) into a more general property-driven training (for any property of choice)~\cite{FischerBDGZV19,SlusarzKDSS23,flinkow2024}.  It is noteworthy that, although robust training by \emph{projected gradient descent}~\cite{goodfellow2015explaining,madry2018towards,kolter2018adversarial} predates verification, contemporary approaches are often related to, or derived from, the corresponding verification methods by optimizing verification-inspired regularization terms. 

The weakest form of property-based training boils down to translating a specification written in a subset of first-order logic into a \emph{loss function}, that serves directly as an optimisation objective within the implementation of a training algorithm. Thus, the training algorithm optimises the neural network to satisfy the desired property.    This translation method is known under the name of \emph{differentiable logic} (or DL) ~\cite{FischerBDGZV19,SlusarzKDSS23,flinkow2024}. Vehicle implements DL as one of its backends~\cite{DBLP:journals/corr/abs-2401-06379} (cf. Fig.~\ref{fig:Vehicle}) and serves as a prototype of a compiler for neural network property specification languages (cf. Fig.~\ref{fig:CV}). Recently, this inspired attempts at formalising different DLs in Coq~\cite{Affeldt24}. 

There are other forms of training for robustness that come with stronger guarantees than DLs, e.g.
 IBP training~\cite{gowal2019scalable,zhang2019stable} and
certified training~\cite{muller2023certified,zhang2020robustness}. However, these usually have limited capacity for property specification; investigation of how these methods may fit into larger verification pipelines is warranted.


\subsection{Other Directions}\label{sec:other}



\subsubsection{Verification of Cyber-Physical Systems.}
When following the diagram of Fig.~\ref{fig:CV}, we did not impose any assumptions on the nature of properties we wish to ensure. In particular, we did not specify whether the ``System'' needs to be a cyber-physical system (CPS). However, CPS with machine learning components is an important safety-critical use case for neural network verification.
%

For example, a neural network may be utilized as a feedback controller for some plant model, typically represented as ordinary differential equations (ODEs) or generalizations thereof like hybrid automata. These are known as \emph{neural networks control systems (NNCS)}.
The introduction of constraints to describe the dynamics of a CPS requires revisiting several blocks of Fig.~\ref{fig:CV}. Specifically, we need to replace the purely symbolic specifications and algorithms with those allowing for continuous variables and differential equations.   

The annual International Competition on Verifying Continuous and Hybrid Systems (ARCH-COMP) 
has a category for this problem class, known as the AI and NNCS (AINNCS) category~\cite{lopez2019archcomp_ainncs,johnson2020archcomp_ainncs,johnson2021archcomp_ainncs,lopez2022archcomp_ainncs,lopez2023archcomp_ainncs}.
Several approaches for addressing the NNCS verification problem have been developed, such as implemented within software tools like CORA~\cite{kochdumper2023nfm}, JuliaReach~\cite{Bogomolov2019}, NNV~\cite{tran2020cav_tool,lopez2023nnv}, OVERT~\cite{sidrane2019safeml}, POLAR~\cite{huang2022atva}, Sherlock~\cite{dutta2018adhs,dutta2019hscc}, ReachNN*~\cite{huang2019reachnn,fan2020reachnn}, VenMAS~\cite{AkintundeBKL20}, and Verisig~\cite{ivanov2019hscc,ivanov2020tecs,ivanov2021cav}.

More broadly, researchers have considered several strategies for the specification of properties of CPS with neural network components~\cite{fremont2019pldi,astorga2023oopsla,calinescu2024tse}.
These cover significant challenges in the CPS domain, ranging from classical software verification problems to real-time systems concerns, scalability, as well as finding suitable specifications~\cite{seshia2022cacm,tran2022mdat,DBLP:journals/corr/abs-2402-10998,DBLP:journals/corr/abs-2405-14058}. Similarly to the standard neural network verification pipeline of Fig.~\ref{fig:CV}, this area would benefit from a more principled programming language support.

%

\subsubsection{Formal Specification of Probabilistic Properties.}
Program synthesis techniques can be valuable allies in producing correct-by-construction software and systems. In particular, the synthesis of logical formulas from a neural network and a dataset (e.g., via Inductive Logical Programming) received long-timed interest~\cite{payani2019inductive}.
Also orthogonal to our work is Probabilistic Programming (as seen in~\cite{manhaeve2021neural}), which aims to provide a language and toolchain to express probabilistic properties of programs. It is clear that neural networks -- seen as programs -- would benefit from a probabilistic specification language. An early example in this direction is the ProbCompCert~\cite{tassarotti2023density} project.
\vspace{-1em}

\subsubsection{Formalisation of Machine Learning.}
So far, research on formalisation of neural networks or optimisation algorithms has developed in isolation from the mainstream neural network verification pipelines summarised in Fig.~\ref{fig:CV}. However, these two lines of research are bound to meet one day. Relevant work in formalisation of machine learning includes: verification of neural networks in Isabelle/HOL \cite{brucker2023verifying} and Imandra~\cite{DesmartinPKD22}; formalisation of piecewise affine activation functions in Coq~\cite{aleksandrov2023formalizing}; providing formal guarantees of the degree to which the trained neural network will generalise to new data in Coq~\cite{bagnall2019certifying}; convergence, in this case of a 
single-layered perceptron in Coq~\cite{murphy2017verified}; verification of neural archetypes in Coq~\cite{de2022use}. The two approaches that came the closest to unifying formalisation and verification in neural network domain are the Vehicle formalisation in Agda~\cite{ADK24} and the formalisation of differentiable logics in Coq~\cite{Affeldt24}, relation of the latter to the verification pipeline of Fig.~\ref{fig:CV} is discussed in Sec.~\ref{sec:ML}.

\section{The Future Roadmap}\label{sec:future}
\newcommand{\tool}[1]{\rotatebox[origin=c]{-90}{#1}}
\begin{table}[t]
\centering
\footnotesize{
\begin{tabular}{ p{4cm} c c c c c c c c | c c } 
\toprule
\textbf{Existing Solutions}
& \multicolumn{2}{c}{\textbf{High-level}} 
& \multicolumn{2}{c}{\textbf{Low-level}}
& \multicolumn{2}{c}{\textbf{Quantised}}
& \multicolumn{2}{c}{\textbf{Software}}
& \multicolumn{2}{c}{\textbf{Future}} \\ 
\cmidrule(lr){1-1}
\cmidrule(lr){2-3}
\cmidrule(lr){4-5}
\cmidrule(lr){6-7}
\cmidrule(lr){8-9}
\cmidrule(lr){10-11}
PL Challenges Addressed
& \tool{Vehicle~\cite{DaggittKKA0SCCL23}} 
& \tool{CAISAR~\cite{girardsatabin2022caisar}} 
& \tool{$\alpha \beta$-Crown~\cite{wang2021beta}} 
& \tool{Marabou~\cite{katz2019marabou}}
& \tool{QEBVerif~\cite{Zhang2023milp}}
& \tool{Aster~\cite{lohar2023milp}}
& \tool{CBMC~\cite{kroening2014cbmc}}
& \tool{ESBMC~\cite{Menezes2024esbmc}}
& \tool{Unified Language}
& \tool{Formal Interfaces}
\\
\midrule
  \S 3.1-3.2. Rigorous Semantics & $\checkmark$ & $\checkmark$ &  &  &  &  & $\checkmark$ & $\checkmark$ & $\checkmark$ & $\checkmark$ \\
  \S 3.3. Embedding Gap      & $\checkmark$ &  &  &  &  &  &  & & $\checkmark$ & $\checkmark$ \\
  \S 3.4. Implementation Gap &  & \checkmark &  & $\checkmark$ & $\checkmark$ & $\checkmark$ & $\checkmark$ & $\checkmark$ & $\checkmark$ & \\
  \S 3.5. Proof Certificates &  &  &  & $\checkmark$ & & & $\checkmark^*$ & & $\checkmark$ & $\checkmark$ \\
  \S 3.6. Supports Training  & $\checkmark$ &  &  &  &  &  &  & & $\checkmark$ & $\checkmark$ \\
\bottomrule
\end{tabular}}
\captionof{table}{\footnotesize{\emph{Examples of existing solutions and the PL challenges they (partially) address. For the sake of variety, we include the existing solutions in four distinct categories:  \textbf{high-level neural network verification DSLs} (Vehicle, CAISAR); best-performing (according to VNNCOMP) \textbf{low-level neural network verifiers} ($\alpha \beta$-Crown, Marabou); \textbf{formalisation and synthesis of quantised neural networks} with mainstream MILP solvers (QEBVerif, Aster); and the use of \textbf{general-purpose software verifiers} (CBMC, ESBMC) in neural-network verification. For the latter, when we mark proof certificate production as $\checkmark^*$, we refer to the generic proof production available for those verifiers, as opposed to the production of the \emph{Farkas witness} for neural network \texttt{UNSAT} problems that is available in Marabou.}}}\label{tab:structure}
\end{table}


The previous section identified five desirable programming language features that neural network verification could benefit from: rigorous semantics, support for handling the embedding and implementation gaps, generation of proof certificates, and rigorous integration of property-driven training into verification pipelines.
 Currently, no single neural network verification tool or framework possesses all five features (see Table~\ref{tab:structure}). 
Moreover, some tools considered leaders in the neural network verification market do not satisfy any. 
The title of this paper reflects our belief that the desirable solution -- 
global specifications that
\emph{formally} explain the expected properties and yield a \emph{formal}
proof that the given implementation respects the specification -- is a challenge in programming language design. In this section, we overview a couple of possible directions that may play a role in future solutions.





\subsection{A Unified Dependently Typed Language}

We believe the idealised solution to be a single language that is expressive enough to implement the machine learning pipeline and, at the same time, encode the desired properties of both the neural networks created by the pipeline and the pipeline itself. 
The following are a non-exhaustive list of the types of properties that should be representable: 
\begin{enumerate}
  \item theoretical results about the convergence of the training process;
  \item correctness of tensor operations that underlie the training;
  \item rich properties of the input data, e.g., constraining inputs to a certain range; 
  \item relation between input data and the weights in the network produced, e.g., robustness;
  \item properties of the floating point numbers being used.
\end{enumerate}
Given the complexity of encoding some of these properties (e.g. numerical stability, robustness), we believe that the expressive power of dependent types is a natural fit. 
We now briefly argue how such a unified dependent-typed language would allow us to make progress towards the challenges outlined in Table~\ref{tab:structure}. 

\begin{enumerate}
\item \textbf{Rigorous semantics.} The meta-theory of dependent types is well studied~\cite{coquand1986coc} so defining rigorous semantics for the language should not be a significant challenge. Furthermore, by implementing all the components in a single language, the friction of aligning the semantics of the different components in the system is significantly reduced.

\item \textbf{Embedding gap.} In a dependently typed language, from one perspective there would be no embedding gap, as any representation changes must be stated explicitly as type conversions. However, from another perspective, working in such a language does not address the fundamental problem of translating the proofs from the problem space to the embedding space. It merely moves the work from the external proofs into the type-conversion functions. Nevertheless, the expressive power of dependent types is more than sufficient to implement the partial solutions proposed by Vehicle (Sec.~\ref{sec:embgap}).

\item \textbf{Implementation gap.} The implementation gap (Sec.~\ref{sec:integrity}) will be resolved, as implementations will respect their types.  For example, consider numerical types. If our specification assumes infinite reals,
there is no way to instantiate an implementation that uses machine floats, as we
will not be able to prove that machine floats is a valid representation of reals.
If our specification is weak and we do not require properties of the operations
or other equalities, then machine floats may be a valid data type for the chosen
constraints.  If we envision implementation to operate on machine floats, we can
describe all the properties of interest (e.g., lack of associativity) in the data
type.  We must understand that we cannot use external libraries such as XLA or
OpenBLAS without verifying them, as this will break all the formal guarantees
in our specification.  Either these libraries have to be verified formally or
we can synthesise the code with similar runtime properties directly from our
specification in a type-preserving way.

\item \textbf{Proof certificates.} In a dependently-typed language, where the specifications are encoded as types, there is no need for separate proof certificates or proof checkers. In particular, the terms themselves act as the proof certificates and the type-checker acts as the proof checker.

\item \textbf{Supports training.} Although at the moment training is carried out in non-dependently typed languages, there is nothing stopping training (e.g.  automatic differentiation or similar algorithms) from being implemented in dependently-typed language. Not only would such an implemention significantly reduce the friction between training and verification, it would also facilitate the integrating property-driven training and verification by viewing it as code synthesis problem.
The key idea here is that generating code from a formal specification is much
easier than checking whether the given code respects the specification. 
\end{enumerate}


\subsection{Formal Interfaces}

However, we are not naive as to the difficulty of implementating such a unified framework. Firstly, it will be an uphill battle to overcome the significant first-mover advantage of existing tools in their respective domains, e.g. training frameworks in Python, C and others and neural network verifiers. Even leaving that aside, work has shown that checking such complex type-based specifications in an efficient manner is still a challenging problem (e.g. Kokke et al.~\cite{KKKAA20}).

Therefore, we believe a more realistic short-term goal is to keep the overarching maximally expressive specification language, but design a compiler that can utilise existing tools to achieve certain subgoals. In particular, we should follow the approach of industry, where the use of many disparate systems is common. In such an environment, the designers of these individual components should not only rigorously pre-define their interfaces, but also provide full formal specifications about their behaviour and, ideally, provide proof certificates that the output satisfies the specification as part of the interface. This would allow the compiler to specify the expected behaviour of a given module and let the programmer choose the best implementation (provided it respects the specification) at that abstraction level.


One possible inspiration for the design of such interfaces could come from the rich literature of behavioural interface specification language (BISL)~\cite{hatcliff2012Behavioral}. A BISL is a family of languages used to specify the expected behaviour of a program at the \emph{function} level, providing a fine-grained level of control on how to precisely describe the function. BISLs usually follow a Hoare triplet-inspired formalism: the programmer should specify the precondition and the control flow; automated provers using weakest-precondition calculus or SAT can then automatically derive preconditions. Drawing from well-known languages like SPARK and Eiffel, it would then become easy to specify invariants on \emph{several functions} at once. Such properties could then be translated back to the original program language (and then checked with the type system) or - if it results in a program that is impossible to represent - checked using external provers. This approach would be representative of what is being done in the industry for critical systems for decades, with JML, Why3~\cite{filliatre2013why3} or SPARK.

%



\section{Conclusion}

We have given some support for our main thesis -- that \emph{neural network verification is increasingly becoming a programming language challenge}. We hope this paper will provoke a stimulating discussion of this topic, helping the programming language community explore the opportunities presented by this new domain. Although we have supported our arguments with references to existing approaches, this is not a survey paper, and we make no claims of bibliographic completeness. 

\section{Acknowledgements}
M. Daggitt and E. Komendantskaya acknowledge the partial support of the EPSRC grant AISEC: AI Secure and Explainable by Construction (EP/T026960/1).
E. Komendantskaya was supported by ARIA: Mathematics for Safe AI grant.  
L. Cordeiro and E. Manino acknowledge the support of the EPSRC grant EnnCore:  End-to-End Conceptual Guarding of Neural Architectures
(EP/T026995/1).
J. Girard-Satabin and Augustin Lemesle were supported by the French Agence Nationale de la Recherche (ANR) grant ANR-23-DEGR-0001 as part of the France 2030 programme.
The work of Isac and Katz was partially funded by the European Union (ERC, VeriDeL, 101112713). Views and opinions expressed are however those of the author(s) only and do not necessarily reflect those of the European Union or the European Research Council Executive Agency. Neither the European Union nor the granting authority can be held responsible for them.

\bibliographystyle{splncs04}
\bibliography{bibliography,references}
\end{document}